# SAP HANA AND ITS PERFORMANCE BENEFITS

By

TIMUR MIRZOEV *          CRAIG BROCKMAN **

*Georgia Southern University*
*\*\* Lockheed Martin*

*ABSTRACT*

*In-memory computing has changed the landscape of database technology. Within the database and technology field, advancements occur over the course of time that has had the capacity to transform some fundamental tenants of the technology and how it is applied. The concept of Database Management Systems (DBMS) was realized in industry during the 1960s, allowing users and developers to use a navigational model to access the data stored by the computers of that day as they grew in speed and capability. This manuscript is specifically examines the SAPHigh Performance Analytics Appliance(HANA) approach, which is one of the commonly used technologies today. Additionally, this manuscript provides the analysis of the first two of the four common main usecases to utilize SAP HANA's in-memory computing database technology. The performance benefits are important factors for DB calculations.Some of the benefits are quantified and the demonstrated by the defined sets of data.*

*Keywords: Database Management Systems (DBMS), SAPHigh Performance Analytics Appliance(HANA),* Structured Query Language (SQL), Relational Database Management system (RDBMS).

## INTRODUCTION

The concept of a Relational Database Management system (RDBMS) came to the fore in the 1970s. This concept was first advanced by Edgar F. Codd in his paper on database construction theory, "A Relational Model of Data for Large Shared Data Banks". The concept of a database table was formed where records of a fixed length would be stored and relations between the tables maintained [1].

The mathematics at the heart of the concept is now known as tuple calculus [2]. The variation on relational algebra served as the basis for a declarative database query language, which in turn formed the basis for the Structured Query Language (SQL). SQL remains as the standard database query language some 30 years later.

Database technology, specifically Relational Database Technology (RDBMS), has seen incremental advancements over recent decades but the competition has narrowed to a few remaining larger entities. The pursuit for improvement has largely left technology practitioners, especially the database administrators, focused on the benefits of performance tuning of the database technology. The infrastructure teams have relied on benefits of the potential compression factors from various RDBMS offerings to help quell the ever expanding footprint of structured and unstructured data that fills the capacity of the typical data center.

As a result, infrastructure teams continually seek hardware refreshes with promises of faster disk performance and improved memory caching to gain new database performance tools. Ultimately,a database administrator is left with only a few tools to improve database performance such as adding and tuning database indexes, which only add to the amount of space required for the database. In the end, the data of concern becomes secondary too and can even become smaller than the indexes themselves, leaving the technology practitioners faced with a diminishing rate of return from their efforts. Technology can only go so far and the physics of spinning disks is reached eventually with the associated costs of competing methods to store, retrieve and query data.

Today, IT professionals are challenged with the task of on going improvements to achieve goals of businesses. Unfortunately, IT budgets do not dynamically grow as fast as business needs. That sequence of events creates majors obstacles for DB infrastructure, deployment, administration and maintenance.

Frequently, IT is not seen as a valuable partner in meeting the objectives of the business and IT is not characterized as the means by which a business can create new opportunities and grow. More difficult times lead to further budget tightening and reductions in the needed hardware refreshes. All of these factors leave no time for innovation, creativity and the lack of budget for evaluating the latest new technology.

The reality is a declining benefit of investment into physical disks. The apparent limitation in the physics of spinning disks leads to questioning previous systems and development paradigms. With the continued need for deeper, mobile and real-time access to all of the companies' data calls for innovative solutions.

Emerging technology in the area of in-memory computing and products like SAP HANA are changing the database technology arena and the analytics field and the overall system architectures [3]. These database advancements can bring improvements in the timeliness of access to data and can enable improved decision making and enable transformation of business processes to make use of this new capability.

## 1. A Historical View

In order to proceed with the subject at hand, an overview of Relational Database Management System (RDBMS) advancements is needed.

The database technology field dates to the 1960s, with the introduction of the Integrated Data Store by Charles Bachman [4]. Prior to 1970 the relational database was conceptual and data was stored in separate files stored on magnetic tape [5]. Memory was expensive, while storage was relatively inexpensive and the Moore's law was a maturing concept without its full validity or its temporal limitations [6].

The proceeding decades of database history can be tabulated into nine distinct eras. In viewing this data alone a slowing trend of advancements can be seen since the mid-1990s. Table 1 outlines the 9 eras.

Through the proceeding decades of database history and computer science theory specifically, Moore's law has proven itself and the linear nature of the number of components in integrated circuits has doubled at a predictable pace. The linear nature of this relationship as it applies to Central Processing Unit (CPU) performance and CPU cost has begun to change very recently. CPU performance improvement has begun to slow and clock rates have remained nearly steady for several years. Meanwhile the cost of these components continues to decline substantially. Furthermore, the cost of memory has also declined substantially. During this same time physical disks have seemed to reach a performance plateau and not offered the same declining cost and improving performance relationship as CPU and RAM have demonstrated.

These changes have enabled hardware advancements including more cores and more CPUs on a given server and far more RAM is now feasible [7]. RAM costs have also changed such that it is less necessary to limit the amount of RAM and use it only for caching of critical processes. Instead, RAM can now go toe-to-toe with fixed disk for storage and also combine with the CPU to provide a tremendous performance differential as compared to disk-based systems and databases [8]. When compared with disk, the physics of spinning platters and their resulting database, read and write characteristics are best measured in milliseconds. RAM arrives without those physical limitations and delivers nanosecond response times [9]. Increasing RAM is one possibility -one must begin to think in terms of terabytes of RAM and no longer gigabytes of RAM, in much the same way computer science has moved beyond kilobytes and bytes of RAM quite some time ago. According to IBM, some applications simply cannot run without in-memory computing even with high-data volumes [10].

| Timeline | Era | Important Products |
|---|---|---|
| Before 1970 | Predatabase | File managers |
| 1970-1980 | Early database | ADABAS, System 2000, Total, IDMS, IMS |
| 1978-1985 | Emergence of the relational model | DB2, Oracle |
| 1982-1992 | Microcomputer DBMS products | dBase-II, R:base, Paradox, Access |
| 1985-2000 | Object-Oriented DBMS | Oracle ODBMS and others |
| 1995-present | Web database | IIS, Apache, PHP, ASP.NET, and Java |
| 1995-present | Open source DBMS product | MySQL, PostgreQL, and others |
| 1998-present | XML and web services | XML, SOAP, WSDL, UDDI, |
| 2009-present | NoSQL movement begins | Apache Cassandra, dbXML MonetDB/XQuery, and others |

Table 1. Outlines the eras in the Timeline of Database Technology Development [5].

In-memory computing does dramatically change the landscape of the database technology [11]. In-memory computing involves the use of main memory not as a cache of memory but instead as a means of storage, retrieval and computational performance in close proximity to the Central Processing Unit (CPU). The use of row and columnar storage of data creates the ability to remove the need for the index which is predominating in the disk-based database world. The lack of the index is a key area of compression and columnar storage allows further compression.

According to IBM and as Shown in Figure 1, "the main memory (RAM) is the fastest storage type that can hold a significant amount of data" [10].

These advancements arrive at a time when RDBMS advancements have increased and the typical IT infrastructure is faced with the declining budgets and the expectations of improving performance results; all at the same time. If a hardware refresh is being planned, Linux-based solutions and in-memory database solutions are increasingly becoming part of the reference architecture. At a minimum, the question of in-memory computing is now being asked as part of architecture planning for infrastructure updates. These technologies are increasingly being accepted as proven and ready for long-term capital investment.

The problem of data growth and the impact it has on run-times and data accessibility is faced by system administrators, programmers and functional analysts. Due to the data growth, a program, system process that used to run each day, can now run once a week. Another problem is the runtime of a report which creates the need for a developer to reduce the user's ability to ask questions of the data, limiting its selection criteria. These problems may also impact system administrators as they are asked to add database indexes to improve the programs performance.

Data growth is viewed as the largest datacenter infrastructure challenge based on Gartner research [12]. Survey data indicated business continuity and data availability were the most critical, followed by cost containment initiatives and maintained or improving user service levels was the third most important factor for data center planning and strategy development [12].

Another way to view the large data analytics is analyze three characteristics of big data – volume, velocity and variety [13].

2. The Setup

The hardware specifications are two-servers were configured to act as one database. When combined the system has eight (8) Westmere EX Intel E7-8870, ten (10) core CPUs (80 cores) with a 2.4 GHz clock rate. The system has two (2) SSD (Solid-State Drives) Fusion-IO drives with 640GB of capacity each for database log volumes. The system has sixteen (16) 600GB 10,000 RPM SAS drives in a RAID-5 configuration for persistence of the data volume in addition to its storage and calculation on sixteen (16) 64GB DDR3 RDIMM.

3. SAP HANA in Action

It has been established that the emerging technology in the area of in-memory computing is making promising advancements and the technology is being applied in ways that were not previously anticipated. According to IBM, SAP High Performance Analytics Appliance (HANA) "enables marketers to analyze and segment their customers in ways that were impossible to realize without the use of in-memory technology" [10]. Some applications are not available without in-memory technology [10].

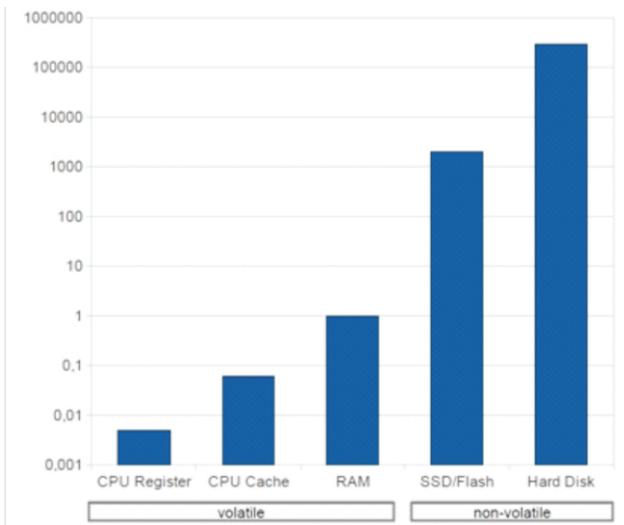

Figure 1. Data Access Times of Various Storage Types, Relative to RAM (logarithmic scale)

Unique application of this emerging technology has been applied within healthcare field, for personalized patient treatment planning in real-time, to greatly accelerated cancer screenings and human genome analysis, out of the box thinking is needed to explore and find the limits of these new technologies [14].

These four use-case areas for SAP HANA are [15]

- real-time replication of data from SAP and non-SAP source systems for business intelligence analytics,

- using this table replication capability to accelerate programs and reports from the SAP ERP system and reading them faster than they can be read from the disk-based database serving the SAP ERP system,

- as a database replacement for SAP Business Warehouse (BW) allowing a full and complete database migration and replacement path from various disk-based technologies to SAP HANA as the in-memory database for SAP Business Warehouse (BW) and

- as a database replacement for SAP ERP allowing a full and complete database migration and replacement path from various disk-based technologies to SAP HANA as the in-memory database for SAP ERP.

When compared to a disk-based database, the in-memory database yields far faster read-rate response times [10], when comparing disk-based technologies the measurements are in milliseconds while in-memory databases are measured in nanoseconds. The unit of measure shift alone is a matter of calculation and not the result of a test to outline the significant change from measurement in 1/1,000 (one one-thousanth) of a second to 1/1,000,000,000 (one one-billionth) of a second.

4. Measured Results

The results were quantified and the system was measured in terms of the aggregation levels of a given report. Typically, a report is severely constrained in selection criteria to aide runtime performance, effectively removing scope from the user's ability to query the data. Another measurement will be in terms of data compression. The source database tables were measured for index and data size and the compression of the data size calculated. As the data is stored in columnar form, the need for the index is removed. Finally, measurements in preparing the data for query including runtime of all background jobs, extracts, transformations, loads, aggregation and index in addition to query time will be compared to the replicated data and its query response time using the in-memory database.

The first of two use-cases accelerated a complex table-set and performed an analytics use-case demonstrating the real-time nature of the data access, the improved response time and the data compression characteristics of an in-memory database. The second use-case utilized the improved read-rate performance to accelerate an existing table-set and performed a database call from an existing program with the table-set residing on SAP HANA, instead of directing the database call to the disk-based database.

The hardware utilized was from one of several hardware vendors partnering with SAP to provide hardware capable of utilizing SAP HANA as an in-memory computing engine.

As discussed previously, disk-based hardware advancements have slowed, the cost of various system components have fundamentally changed and the user base along with the system administrator is faced with an ever increasing growth of data size and along with increasingly complex data with the result being slower and slower system performance.

The first of two use-cases accelerated a complex table-set and performed an analytics use-case demonstrating the real-time nature of the data access, the improved response time and the data compression characteristics of an in-memory database.

In this first use-case the SAP HANA appliance hardware was installed and configured and the necessary replication of the SAP ERP source tables was established. The initial load of data took as much as 120 hours due to CPU and RAM restrictions on an intermediate server. Additional tuning and increased allocation of CPU and RAM on that intermediate server allowed subsequent loads to be completed on the same dataset in less than fourteen hours.

With this real-time replication established and these initial one-time loads, these datasets are now continually updated as the source database is changed, in real-time.

After some data modeling of these tables and development of a set of reports, query of the dataset using the SAP HANA in-memory platform was tested. Previously, the dataset was accessible as a static result running each night. As is typical the dataset grew and as such run-time began to exceed the daily iteration of assembling the data for analysis. With the data now only accessible weekly, a user had less access to the necessary data to run the business. The combination of background programs and data extracts to an external data warehouse system combined for a total processing time of 1,131 minutes or 18.9 hours. The data set was comprised of approximately 40 tables, the largest of which was over 700 Million rows in size.

As indicated by Figure 2, the prior dataset contained aggregations limiting the analysis scope. The user had limited access to the data as constraints were placed on the selection criteria in order to aid run-time performance. Using SAP HANA these constraints are removed and the user can now query the full dataset range.

As for data compression, the data set of approximately 40 tables on the disk-based database utilized 765 GB of total space and more than half of this space was attributed to index space (428.2 GB) while the data space accounted for only 336.3 GB. The 336.3 GB replicated to HANA now accounts for 108.2 GB of compressed data in memory. The index space is not replicated to SAP HANA as the indexes are not necessary for the in-memory database.

As noted by Figure 2, the total processing time of 1,131 minutes or 18.9 hours, which is the combination of background programs and data extracts to an external data warehouse system including the query response of the static report is now compared to the response time of less than 1 minute for the dynamic report. The dynamic report is now available in real-time, all day, every day and the user no longer has to wait each week to receive an aggregated, static report. This is a significant advantage over the previous approaches that did not utilize in-memory technology.

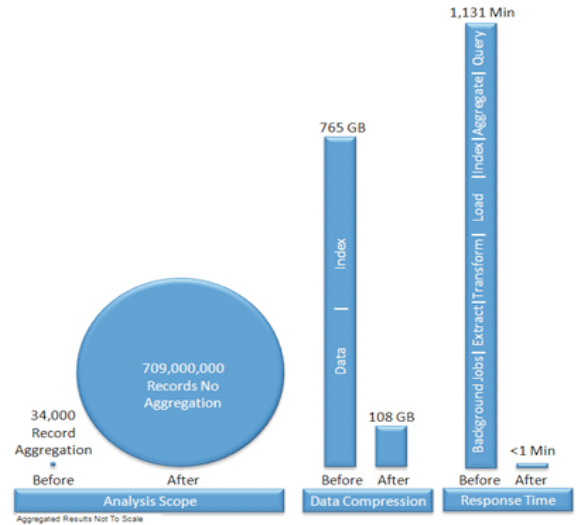

Figure 2. Comparison of Changes in Aggregation Level, Data Compression and Data Processing

The second use-case utilized the improved read-rate performance to accelerate an existing table-set and performed a database call from an existing program with the table-set residing on SAP HANA instead of directing the database call to the disk-based database.

The secondary database connection to SAP HANA facilitated an accelerated read from existing standard and customized code. The accelerated read-rate performance yielded significant improvements even on unoptimized code.

As specified, the system had two (2) SSD Fusion-IO drives with 640GB of capacity each for database log volumes and sixteen (16) 600GB 10,000 RPM SAS drives in a RAID-5 configuration for persistence of the data volume in addition to its storage and calculation on sixteen (16) 64GB DDR3 RDIMM. Support for a Storage Area Network (SAN) is not required as on-device persistence is part of the solution; other back-up solutions are supported.

Table 2 presents the data compression summarized. Dependent on the cardinality of the table, the data compression for a given table ranged from <1x compression to near 6x compression; overall this data set compressed 3x. When comparing the compressed data set in-memory to the data set including indexes on the traditional database the data compression for a given table may range from <1x compression to over 13x compression, overall this data set compressed 7x.

As outlined by Figure 3, the unoptimized code calling the traditional database had a run-time of 291 seconds. The same unoptimized code calling the same table set replicated on SAP HANA had a run-time of 22 seconds. The difference represented a 13x improvement in the run-time of the sample program tested. Identifying unoptimized code and mitigation of the unoptimized code further improved the run-time on the traditional database to 12 seconds, an improvement of 24x versus the baseline. The now optimized code calling SAP HANA now has a run time performance of .67 seconds, an improvement of 434x versus the baseline.

| # | TABLE | Index Size/KB | Table Size/KB | Source DB Count | HANA DB Count | HANA DB Size/KB | Compression w/o indexes | Compression with indexes |
|---|---|---|---|---|---|---|---|---|
| 1 | CATSDB | 28,073,600 | 28,499,648 | 51,782,160 | 51,782,160 | 10,767,235 | 2.65 | 5.25 |
| 2 | CEPC | 128 | 384 | 1,499 | 1,499 | 309 | 1.24 | 1.66 |
| 3 | CEPCT | 256 | 192 | 1,551 | 1,551 | 317 | 0.61 | 1.41 |
| 4 | COBK | 23,347,200 | 15,616,000 | 73,342,100 | 73,342,100 | 2,977,877 | 5.24 | 13.08 |
| 5 | COEP | 376,338,432 | 280,985,600 | 709,312,238 | 709,312,238 | 96,832,516 | 2.90 | 6.79 |
| 6 | COSP | 7,536,640 | 9,437,184 | 23,321,576 | 23,321,576 | 1,701,459 | 5.55 | 9.98 |
| 7 | COSS | 13,107,200 | 17,152,000 | 34,249,722 | 34,253,894 | 397,967 | | |
| 8 | CSKA | 4,096 | 3,072 | 25,300 | 25,300 | 1,941 | 1.58 | 3.69 |
| 9 | CSKS | 1,024 | 2,048 | 6,285 | 6,285 | 1,034 | 1.98 | 2.97 |
| 10 | CSKT | 896 | 768 | 6,336 | 6,336 | 1,332 | 0.58 | 1.25 |
| 11 | CSKU | 6,144 | 416 | 36,639 | 36,639 | 4,010 | 0.10 | 1.64 |
| 12 | CSLA | 256 | 512 | 3,782 | 3,782 | 379 | 1.35 | 2.03 |
| 13 | CSLT | 576 | 448 | 3,785 | 3,785 | 573 | 0.78 | 1.79 |
| 14 | PA0000 | 3,072 | 5,120 | 38,196 | 38,196 | 5,430 | 0.94 | 1.51 |
| 15 | PA0001 | 28,672 | 10,240 | 38,197 | 38,197 | 7,113 | 1.44 | 5.47 |
| 16 | PA0002 | 14,400 | 10,240 | 38,197 | 38,197 | 6,531 | 1.57 | 3.77 |
| 17 | PA0315 | 5,120 | 6,144 | 38,197 | 38,197 | 5,554 | 1.11 | 2.03 |
| 18 | PRHI | 79,104 | 82,880 | 1,094,253 | 1,094,253 | 221,160 | 0.37 | 0.73 |
| 19 | PROJ | 16,384 | 27,392 | 41,683 | 41,683 | 12,110 | 2.26 | 3.61 |
| 20 | PRPS | 360,448 | 737,280 | 1,094,253 | 1,094,253 | 449,559 | 1.64 | 2.44 |
| 21 | SETHEADER | 7,168 | 6,144 | 41,064 | 41,064 | 3,765 | 1.63 | 3.54 |
| 22 | SETHEADERT | 7,168 | 9,216 | 140,628 | 140,628 | 18,416 | 0.50 | 0.89 |
| 23 | SETLEAF | 27,648 | 33,792 | 314,009 | 314,009 | 36,733 | 0.92 | 1.67 |
| 24 | SETLINET | 4,096 | 4,224 | 44,033 | 44,033 | 5,106 | 0.83 | 1.63 |
| 25 | SETNODE | 6,144 | 3,072 | 38,745 | 38,745 | 4,156 | 0.74 | 2.22 |
| 26 | T000 | 64 | 64 | 4 | 4 | 36 | 1.78 | 3.56 |
| 27 | T001 | 64 | 128 | 246 | 246 | 121 | 1.06 | 1.59 |
| 28 | T001W | 64 | 64 | 92 | 92 | 186 | 0.34 | 0.69 |
| 29 | T009 | 64 | 64 | 13 | 37 | 42 | 1.52 | 3.05 |
| 30 | T009B | 384 | 384 | 1,235 | 1,641 | 168 | 2.29 | 4.57 |
| 31 | TKA01 | 64 | 64 | 97 | 97 | 174 | 0.37 | 0.74 |
| 32 | TKA02 | 64 | 64 | 175 | 175 | 55 | 1.16 | 2.33 |
| 33 | ZACRNYM | 64 | 64 | 205 | 205 | 30 | 2.13 | 4.27 |
| 34 | ZACRNYM_C | 64 | 64 | 58 | 58 | 41 | 1.56 | 3.12 |
| 35 | ZALTDEPT | 64 | 64 | 264 | 264 | 72 | 0.89 | 1.78 |
| 36 | ZJOBCODE | 576 | 1,024 | 17,032 | 17,032 | 1,485 | 0.69 | 1.08 |
| | Totals (KB) | 448,977,408 | 352,636,064 | 895,073,849 | 895,078,451 | 113,464,992 | 3.11 | 7.06 |
| | Totals (GB) | 428.18 | 336.30 | | | 108.21 | | |

Table 2. Comparison of Data Compression by Table with and without Indexes

## 5. Conclusions and the Future of the In-Memory Computing

In-memory computing is making promising advancements and the technology is being applied in ways that were not previously anticipated. The findings of this manuscript support and outline the performance results on two specific data sets within the context of two of the four primary use-cases for SAP HANA.

This manuscript's limitations included testing only one type of software by SAP - High Performance Analytics Appliance (HANA). Additionally, the factors such as loss of data and the risks and costs associated with the SAP HANA technology were beyond the scope of this study. Hence the following recommendations can be made for future research

- A comparative study of in-memory technologies would be desirable to further analyze the pros and cons of in-memory computing

- Financial implications and risks should be examined for in-memory applications

- As accountability for lost data keeps increasing, data loss and vaulting of cache memory for in-memory applications needs to be analyzed

- The limitations of RAM and data sets for in-memory applications need to be further examined

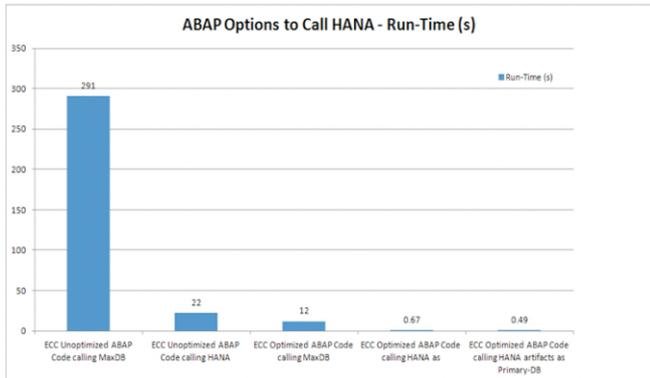

Figure 3. Comparison of Sample Program Response Times on Traditional Database and SAP HANA with and without Code Optimization.

- Another recommendation is to consider in-memory computing and specifically SAP HANA as an emerging technology and to evaluate the integration of this new technology into the architecture where high database input and output demand exists. Specifically, for analytics and acceleration of existing standard and custom ABAP code where the high database input and output is predominately database read activities.

This manuscript addressed only certain use cases of the SAP HANA technologies.

The first use-case outlined the ability to take untimely access to static data sets and enable real-time access to dynamic data sets. The change in analysis scope and timeliness in access to data can enable improved decision making and enable transformation of business processes to make use of this new capability.

The second use-case outlined the ability to access the SAP HANA database through a secondary database connection, and utilizing existing SAP ERP ABAP code to access replicated tables in the in-memory database. The program can yield significant improvements in run-time even on unoptimized code.

As the traditional disk-based architecture data footprint grows at over 36% per year [16], innovations such as in-memory computing are increasing in adoption rate and organizations adopting this technology are also gaining a competitive business advantage [10].

When hardware refreshes are planned, SAP HANA and Linux based operational systems require a thorough review for integration into the architecture and where business requirements demand dynamic access to increasing complex data sets in real-time. Real-time terminology is the key in understanding the benefits of in-memory computing. Some applications and techniques would not be even possible without in-memory technologies [10].

"In May 2011, the market research firm IDC released its Digital Universe report estimating data growth. The IDC researchers estimated that stored data grew by 62 percent or 800,000 petabytes in 2010. By the end of 2011, 1.2 zettabytes (ZB) of information will be held in computer storage systems, according to IDC.

And that is only the beginning of the story. By 2020, data will have grown another 44-fold. Only 25 percent of that data will be original. The rest will be copies [17]. In a survey of 237 respondents from a broad range of organizations 12% indicated they had installed an in-memory database in the last 5 years, 27% between 2 and 5 years ago, 24% between 1-2 years ago, 15% within the last 6 months to 1 year ago and 22% within the last 6 months. The data indicates the technology has had some attention for nearly 5 years, and while it is still maturing to its potential it is certainly a technology worth further consideration whether it has been evaluated previously or not.

Adding to the problem of exploding data is the new reality of a declining benefit of investment on physical disks and an apparent limitation in the physics of spinning disks. With the continued need for real-time mobile access to all of a companies' data, this convergence technology is creating the demand for innovative solutions such as in-memory computing.

[17]. King, E. (2011). The Growth and Expanding Application of In-Memory Databases Retrieved from http://www.loyola.edu/lattanze/workingpapers/WP0611-107.pdf

## ABOUT THE AUTHORS

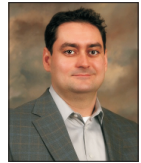

*Dr. Timur Mirzoev is currently working as a Professor of Information Technology Department at Georgia Southern University, College of Engineering and Information Technology. He received his Ph.D. in Technology Management from Indiana State University in 2007. He holds an MBA, an MS and two Bachelor's Degrees. He heads the International VMware IT Academy Center and EMC Academic Alliance at Georgia Southern University. His research interests include Cloud Computing, Virtualization, Storage Areas in which he is commercially certified.*

*Craig Brockman is a business analyst for Lockheed Martin. He has over 14 years of experience in aeronautics, IT Project Management, IT systems strategies, RFID implementations, programming and many other technologies.*